\documentclass[aps,twocolumn,showpacs,prb]{revtex4-1}
\usepackage[english]{babel}
\usepackage{amsmath}
\usepackage{graphicx} 
\usepackage{times}
\usepackage{epsfig} 

\usepackage{color}
\definecolor{nred}   {RGB}{224,0,0}
\definecolor{nblue}  {RGB}{28,130,185}
\definecolor{dgreen} {RGB}{78,138,21}
\definecolor{norange}{RGB}{230,120,20}

\begin{document} 

\title{Dynamical conductivity and its fluctuations along the crossover to many-body localization}

\author{Osor S. Bari\v si\' c$^1$, Jure Kokalj$^{2,3}$, Ivan Balog$^1$, and Peter Prelov\v{s}ek$^{2,4,5}$}
\affiliation{$^1$Institute of Physics, HR-1000 Zagreb, Croatia}
\affiliation{$^2$Jozef Stefan Institute, SI-1000 Ljubljana, Slovenia}
\affiliation{$^3$ Faculty of Civil and Geodetic Engeneering, University of
Ljubljana, SI-1000 Ljubljana, Slovenia}
\affiliation{$^4$ Faculty of Mathematics and Physics, University of
Ljubljana, SI-1000 Ljubljana, Slovenia}
\affiliation{$^5$Arnold Sommerfeld Center for Theoretical Physics, Ludwig-Maximilians-Universit\" at  M\" unchen, D-80333 M\" unchen, Germany }


\begin{abstract}

We present a numerical study of the many-body localization
  (MBL) phenomenon in the high-temperature limit within an anisotropic Heisenberg model with random local fields. Taking the dynamical spin conductivity $\sigma(\omega)$ as the test quantity, we investigate the full frequency dependence of sample-to-sample fluctuations and their scaling properties as a function of the system size $L\leq 28$ and the frequency resolution. We identify differences between the general interacting case $\Delta>0$ and the anisotropy $\Delta=0$, the latter corresponding to the standard Anderson localization. Except for the extreme MBL case when the relative sample-to-sample fluctuations became large, numerical results allow for the extraction of the low-$\omega$ dependence of the conductivity. Results for the d.c. value $\sigma_0$ indicate a crossover into the MBL regime, i.e. an exponential-like variation with the disorder strength $W$. For the same regime, our numerical analysis indicates that the low-frequency exponent $\alpha$ exhibits a small departure from $\alpha\sim 1$ only.

\end{abstract}


\maketitle

\section{Introduction}

The phenomenon of many-body localization (MBL) has been originally suggested for weak disorders,\cite{fleishman80,basko06} 
arguing that interacting systems may exhibit a mobility edge separating low-energy many-body (MB) localized states 
from delocalized ones. In this respect, there is a clear analogy with the single particle spectrum that characterizes the Anderson localization.\cite{anderson58,mott68,kramer93} For the localization in noninteracting (NI) systems the essential ingredient is the phase 
coherence of single-particle states. 
However, the latter is lost in an interacting system due to the scattering among particles.
This makes a proof of the existence of the mobility edge difficult for models with interaction and most studies concentrate on the large disorder limit for which all the MB states are expected to be localized (for a recent overview and references see, e.g., Ref. \onlinecite{gopal15}). 

Numerical solutions of finite one-dimensional (1D) MB quantum systems \cite{oganesyan07,monthus10,berkelbach10} indicate
that in interacting fermion systems strong disorder can effectively induce the MBL phase. The latter has been characterized by several novel features: a) The absence of the d.c. transport 
(and in particular of the d.c. conductivity) at any temperature $T$,\cite{berkelbach10,agarwal15,gopal15}
b) the Poissonian level statistics in contrast to the Wigner-Dyson one in generic interacting systems,\cite{oganesyan07,pal10,lev15,luitz15}
c) generally nonergodic behavior\cite{pal10,huse13,lev14,torres15,johri15} and the existence of conserved local quantities,\cite{serbyn13,huse14} 
d) a discontinuity in the one-particle occupation,\cite{bera15} and e) a very slow (logarithmic) growth of the entanglement entropy\cite{znidaric08,bardarson12,vosk13,vosk15,luitz16} as well as of the energy upon driving,\cite{levi15,kozarzewski16} 
and f) possible subdiffusive behavior on the delocalized side.\cite{agarwal15,gopal15,luitz16,luitz116,znidaric16}
Besides the theoretical curiosity there are also experimentally
relevant MBL systems, in particular cold atoms in optical lattices\cite{schreiber15,bordia16,kondov15} or in real materials, e.g., 
modeled by random spin chains \cite{herbrych13}. For our study of transport properties a particular reference is the experiment 
on a disordered cold-fermion system on optical lattice, driven by an external force \cite{kondov15}, since the quantity measured 
(steady velocity)  should correspond  closely to  the d.c. conductivity $\sigma_0$, discussed in this paper.

While the above characteristics appear to be established deep inside the MBL phase, there evidently remain open questions. In particular, 
it is a challenge to establish whether the transition between the (normal) ergodic and the MBL regime is a well-defined phase transition\cite{basko06,monthus10,huse14,vosk13,vosk15,serbyn15,lev15,potter15} with a possible mobility edge in 
the energy (temperature) spectrum,\cite{basko06,luitz15} or merely a
crossover,\cite{barisic10,Carleo13}  although presumably quite
  a sharp one.  

Closely related is the proper understanding of fluctuations of relevant observables when evaluated for finite systems. 
While due to random nature of disorder the Gaussian fluctuations are expected in ergodic systems, 
anomalous Griffiths-like statistical distribution has been claimed within the regime between the ergodic and the MBL phase.\cite{agarwal15,gopal15} 
It is quite plausible that such statistical properties also affect the meaning of calculated dynamical quantities. In the context of the MBL the most interesting is the d.c. value of dynamical conductivity $\sigma_0=\sigma(\omega \to 0)$ and the low-frequency behavior,

\begin{equation}
\sigma(\omega) \sim \sigma_0 + \zeta |\omega|^\alpha, \label{sig}
\end{equation}

\noindent where $\alpha\leq2$ is a nontrivial exponent discussed in several studies.\cite{agarwal15,gopal15,steinigeweg15} 
Our aim is to clarify properties of $\sigma(\omega)$ in the intermediate regime between the ergodic and the MBL phase and 
for this purpose also to investigate sample-to-sample (STS) fluctuations of $\sigma(\omega)$ in 
the search for anomalous behavior indicating distinct phases.

The paper is organized as follows: In Sec.~II we present the model and the applied numerical methods.
In Sec.~III we present general features of calculated $\sigma(\omega)$ and concentrate on the analysis of 
fluctuations (STS variations) of dynamical spectra, and in particular on the fundamental difference between the 
interacting ($\Delta>0$) and the noninteracting ($\Delta=0$) systems. In the following Sec.~IV we display 
the behavior of the sample-averaged dynamical conductivity, in particular of the most challenging low-$\omega$ 
regime. Conclusions and implications are presented in Sec.~V.

\section{Model and numerical methods}

As the MBL prototype model we consider the 1D anisotropic Heisenberg model with random local fields,
\cite{znidaric08,karahalios09,agarwal15,gopal15,steinigeweg15}

\begin{equation}
H = J \sum_{i} [ \frac{1}{2} ( S^+_{i+1} S^-_i + S^-_{i+1} S^+_i) + \Delta   S^z_{i+1} S^z_i ]  +
\sum_i h_i S_i^z\;.
\label{ahm}
\end{equation}

\noindent Periodic boundary conditions are assumed and $J=1$ is used as the unit of energy. The model (\ref{ahm}) is 
the 1D equivalent to the $t$-$V$ model of interacting spinless fermions with random onsite
energies $h_i$, investigated by a number of authors.\cite{oganesyan07,monthus10,berkelbach10,
barisic10,lev15} For $h_i$ we take the uniform probability distribution $P(|h_i|<W/2)=1/W$, standard in most studies. 

As the quantity of interest we choose the high-temperature ($T \gg J$) dynamical (spin) conductivity $\sigma(\omega)$, expressed as 

\begin{eqnarray}
\sigma(\omega)&=& T \tilde \sigma(\omega) =
\frac{1}{L } \mathrm{Re} \int_0^\infty dt~e^{i \omega^+ t}  \langle j(t) j(0) \rangle = \nonumber \\
&=& \frac{\pi }{L N_{st}} \sum_{n \neq m}  |\langle n| j| m\rangle|^2 
\delta(\omega - \epsilon_m +  \epsilon_n),\label{eq04}
\end{eqnarray}

\noindent where $j= (i J/2)  \sum_{i} ( S^+_{i+1} S^-_i - S^-_{i+1} S^+_i)$ is the (spin) current operator and 
$N_{st}$ the number of MB states.  For calculations of the sample-averaged $\sigma(\omega)$ and 
its STS fluctuations
we use two methods based on the exact diagonalization (ED). In both of them we restrict 
our analysis to the system without a uniform magnetic field or $S^z_{tot}=0$.
The first one is the full ED allowing up to $N_{st} \sim 10^4$ for each of $N_d\sim 100$ samples with random $h_i$, reaching 
$L\leq 16$. In the special case $\Delta=0$, Eq.~(\ref{ahm}) transforms into the (Anderson) model of 
NI spinless fermions,\cite{anderson58,mott68,kramer93} which is solved here by the ED within the single-particle basis for large systems, typically $L\sim 1.6\times10^4$. 

The majority of the results are obtained via the microcanonical-Lanczos method (MCLM),\cite{long03,karahalios09,barisic10} best suited for dynamical quantities at elevated $T > 1$. Its computational requirements are essentially equivalent to the ones for the ground-state Lanczos ED, but with an increased number of Lanczos steps $M$, in order to improve the frequency resolution $\delta \omega\sim L/M$. We are able to obtain results for $L=28$, $N_{st} \sim 4\times10^7$ and $M\sim 10^4$, with typical $\delta\omega \sim 2\times10^{-3} > \delta\epsilon$. $\delta 
\epsilon$ is the characteristic MB level separation (e.g., $\delta\epsilon \sim 10^{-3}$ for $L=16$,
and $\delta\epsilon \sim 10^{-6}$ for $L=28$). 
Spectra are broadened with Gaussians characterized by the frequency width $\eta$. The calculated $\sigma(\omega)$  
has a macroscopic meaning providing $\eta > \delta \epsilon$, while for smaller $\eta$ results involve finite-size and level-statistics effects.  
E.g., for $\omega\lesssim\delta\epsilon$, any level repulsion necessarily affects frequency dependencies of $\sigma(\omega)$ and STS fluctuations.

\section{General features and fluctuations of the spectra}\label{SecFluct}

\begin{figure}[tbh]
\includegraphics[width=86mm]{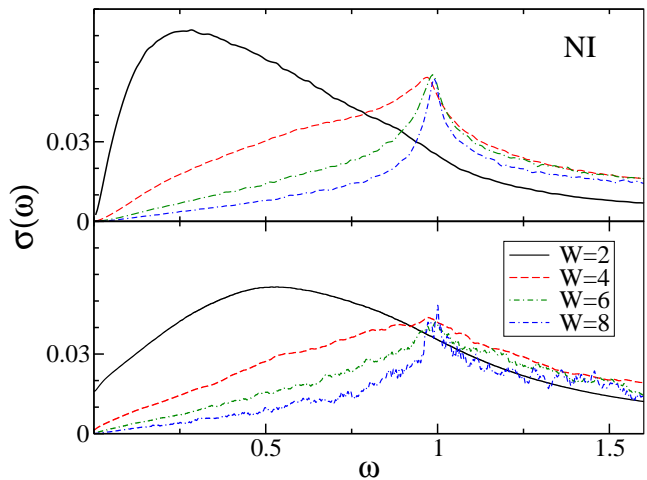}
\caption{(Color online) Large-$T$ dynamical conductivity $\sigma(\omega)$ 
for disorders $W=2-8$ for two cases:
a) $\Delta=0$ (Anderson) model evaluated on a chain with $L=16000$ sites, b) interacting 
$\Delta=1$ case, calculated for $L=28$ using MCLM ($\eta=0.003$).}
\label{fig1}
\end{figure}

Before discussing more delicate issues, we present in Fig.~\ref{fig1} gross results of sample-averaged 
$\sigma(\omega)$ for the NI ($\Delta=0$) and the interacting $\Delta=1$ case, respectively, for various
disorders $W=2-8$, with $\eta \sim 4\times 10^{-3}$. For large
$W$ we note that the general features are very similar in both cases,\cite{barisic10,gopal15,steinigeweg15} e.g.,  
the locations of the maxima are at $\omega\sim1$. 
Essential differences occur for low $\omega \ll 1$. While
for $\Delta=0$ there is a clear drop towards $\sigma_0 = 0$ for all $W$, for $\Delta=1$ we find a rather broad regime 
in which $\sigma(\omega)$ follows the low-$\omega$ behavior in Eq.~(\ref{sig}), with 
$\sigma_0>0$ and $\alpha\sim1$.\cite{karahalios09,barisic10} This behavior will be elaborated further on.

\begin{figure}[tbh]
\includegraphics[width=86mm]{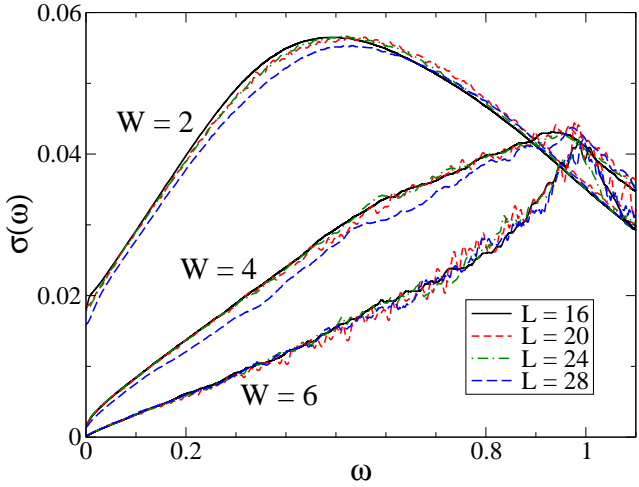}
\caption{(Color online) $\sigma(\omega)$ compared for different system sizes $L$ and three values of $W$
at fixed $\Delta=1$.  $L=16$ system is calculated via the ED, whereas for $L=20,24,28$ we employ MCLM .}
\label{fig2}
\end{figure}

In order to estimate the possible influence of finite-size
  effects, we present in Fig.~2 the direct comparison of the results
  for $\sigma(\omega)$ for fixed $\Delta=1$ but various $W=2,4,6$, as
  obtained for different sizes $L=16 - 28$.  Here, for $L=16$ we use
  the ED, while for larger $L= 20 - 28$ we use the MCLM.  It is rather
  obvious that deviations are hardly visible (taking into the account also
  that for $L=28$ much smaller sampling $N_d = 16$ was used). In
  particular, no systematic trend can be recognized either at high
  $\omega \sim 1$ or low $\omega < 0.1$.  It is, however, not excluded
  that there might be some peculiar behavior below our $\omega$
  resolution, i.e., in the regime $\sigma(\omega <\eta \sim 0.003)$.

To validate the interpretation of sample-averaged $\sigma(\omega)$, we discuss first its relative STS fluctuations,

\begin{equation}
 r_{\eta}(\omega)= 
\sqrt{\langle (\sigma^k_\eta(\omega) - \sigma(\omega) )^2\rangle}/\sigma(\omega), \label{reta}
\end{equation}

\noindent where $\sigma^k_\eta(\omega)$ is a response of a single disorder realization $k$. 
$\sigma(\omega)=\langle\sigma^k_\eta(\omega) \rangle$ are sample-averaged spectra, shown in Figs.~(\ref{fig1}) and (\ref{fig2}), which are essentially $\eta$ and $L$ independent for 
$L\geq16$. Still, by varying $\eta$ and $L$ and by calculating $r_{\eta}(\omega)$ important information on the MBL physics can be obtained.

\begin{figure}
\includegraphics[width=86mm]{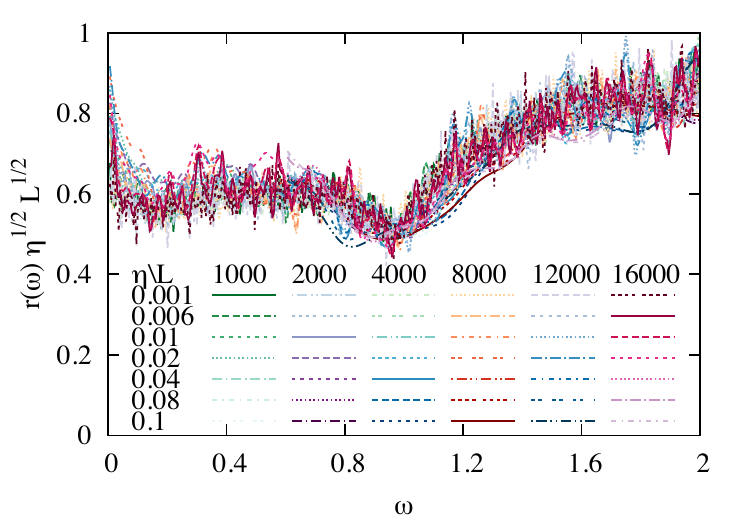}
\caption{(Color online)  Scaled fluctuations of dynamical conductivity 
$r_{\eta}(\omega)\sqrt{\eta\;L}$ for the NI disordered (Anderson) model $W=4$. Here we omit some large $\eta$ results at 
low $\omega$ which deviate from scaling due to stronger $\omega$ dependence.}
\label{fig3}
\end{figure}

Let us consider a coarse-grained description of spectra
$\sigma^k_\eta(\omega)$, where $\eta$ characterizes the frequency
bin. Using the definition Eq.~(\ref{reta}) for $r_\eta$, we presume
that $r_\eta$ is a slowly varying function within the frequency scale
of our interest, $\delta\epsilon\ll\eta\ll 1$. For values of
$\sigma^k_\eta(\omega)$ that are fully random between neighboring
bins, the lack of correlation leads to a simple scaling of $r_\eta$
upon increasing the bin-width $\eta\rightarrow n\eta$ and $r_{n\eta}=
r_\eta/\sqrt n$. Additionally, localization divides the system into
$K=L/l_{NI}^*$  independent sections, and therefore the  contribution
to each bin in Eq.~(\ref{eq04}) is given by $K$  independent variables
(spectra), which directly yields self-averaging, i.e.
$r_{\eta}\propto1/\sqrt{K}\propto1/\sqrt{L}$. This is the behavior
observed in Fig.~\ref{fig3} for the NI system:
$r_{\eta}=b(\omega)/\sqrt{\eta L}$ over multiple scales of $\eta$ and
$L$, with $b(\omega)\sim1$ being model parameter dependent
only. 

\begin{figure}[t]
\includegraphics[width=86mm]{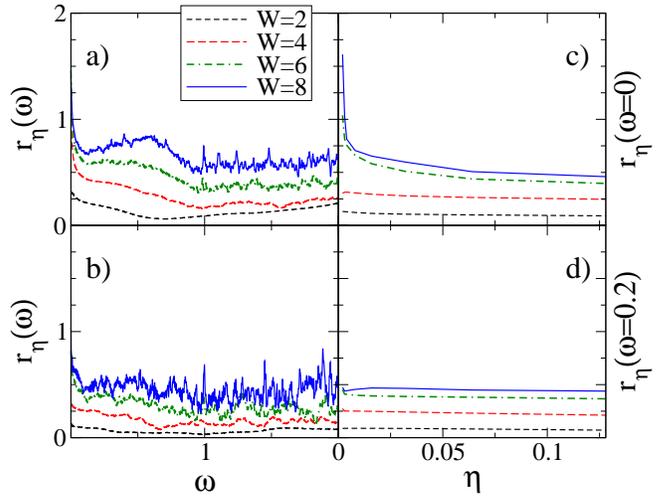}
\caption{(Color online) Fluctuations $r_\eta(\omega)$ for $\Delta=1$ and different $W$: 
a) $r_\eta(\omega)$ obtained by ED for $L=16$ and b) by MCLM for $L=28$,
both at fixed $\eta=0.003$. Panels c) and d) show $\eta$-dependence of 
$r_\eta(\omega)$, at fixed $\omega=0,\;0.2$, respectively, as obtained by MCLM.}
\label{fig4}
\end{figure}

Spectral correlations spreading over $n_\xi$ neighboring bins change
the scaling properties of $r_\eta$. In particular,  
upon increasing the bin-width $\eta\rightarrow n\eta$,
one gets $r_{n\eta}\propto\sqrt{n_\xi/n}\;r_\eta$.
Furthermore, it is clear that for $n_\xi\gtrsim n$ all values of $\sigma^k_\eta(\omega)$ within the large bin $n\eta$ remain correlated, 
behaving as a single random variable, $r_{n\eta}\sim r_\eta$. In connection to the latter behavior, we turn our discussion 
to properties of $r_\eta(\omega)$ for interacting systems for $\Delta=1$, shown in Fig.~\ref{fig4}. 
As seen in Fig.~\ref{fig4}b,c (with the exception of 
$\omega\rightarrow0$ behavior for $W>6$) the interacting case is (similarly as the NI case)
characterized by a weak frequency dependence of $r_\eta(\omega)$.
However, in contrast to the NI case in Fig.~\ref{fig3}, one observes in Fig.~\ref{fig4}c,d a fundamental 
difference in the scaling behavior as a function of $\eta$. In particular, as shown in Fig.~\ref{fig4}d for finite $\omega=0.2$,  
$r_{\eta}$ does not exhibit any significant dependence on $\eta$ over multiple scales, $0.002\leq\eta\leq 0.1$.
This is in part the case for $\omega\sim0$ in Fig.~\ref{fig4}c as well. The exception is the MBL regime $W>6$, for which fluctuations $r_\eta(0)$ become larger due to a very small (or vanishing) sample-averaged value of $\sigma_0$.

At present we cannot give a detailed explanation for the spectral
correlations, emerging for the interacting $\Delta>0$ case 
and leading to correlated bins $n_\xi \gg 1$. Still, an evident argument regarding the role of interaction $\Delta>0$ 
can be given  in terms of frequency moments of 
$\sigma(\omega)$. For $T\to \infty$, assuming the grandcanonical distribution these moments, $m_{2l} =\int_{-\infty}^{\infty} \omega^{2l}\sigma(\omega) d\omega$, 
can be obtained analytically for $L\to\infty$. In particular, one obtains for an arbitrary configuration of $h_i$,

\begin{equation}
m_0=\frac{J^2}{8}, \qquad m_2=\frac{J^2}{16} (J^2 \Delta^2 + \frac{4}{L}\sum_i h_i^2). \label{m02}
\end{equation}
While $m_0$ is independent of disorder, $m_2$ involves STS fluctuations. The cumulant
$C_2 = m_2/m_0$ and its normalized fluctuations are given by

\begin{equation}
\bar C_2 = \frac{1}{2} J^2 \Delta^2 \left[ 1 + \frac{w^2}{6}\right] , \quad
\frac{\delta(C_2)}{\bar C_2 }= \frac{1}{ \sqrt{5 L}} \frac{2w^2}{3 + w^2}, \label{vardel}
\end{equation}

\noindent where $w=W/(J \Delta)$. For large disorder $w \gg1$, the contribution to the fluctuations in Eq.~(\ref{vardel}), 
associated with the interaction $\Delta$, becomes small. It is plausible that higher cumulants $C_{2l}$ have similar 
behavior. The crucial difference is that for the NI ($w=\infty$) systems there is no dimensionless parameter which would control fluctuations in Eq.~(\ref{vardel}), justifying the NI scaling as shown in Fig.~\ref{fig3}. On the other hand, for $\Delta \neq 0$ a new finite frequency scale $\Delta\omega>0$ sets in, which is determined by $w$ and represents the frequency correlations in single-sample $\sigma^k_\eta(\omega)$, leading to $\eta$-independent spectra for $\eta<\Delta \omega$. In fact, this lack of $\eta$ dependence is well visible in our numerical results for individual interacting $\Delta>0$ spectra $\sigma^k_\eta(\omega)$ corresponding to different disorder realizations. However, the actual form of $\Delta\omega(w)$ remains to be understood.

\section{Averaged dynamical conductivity}

\begin{figure}[t]
\includegraphics[width=86mm]{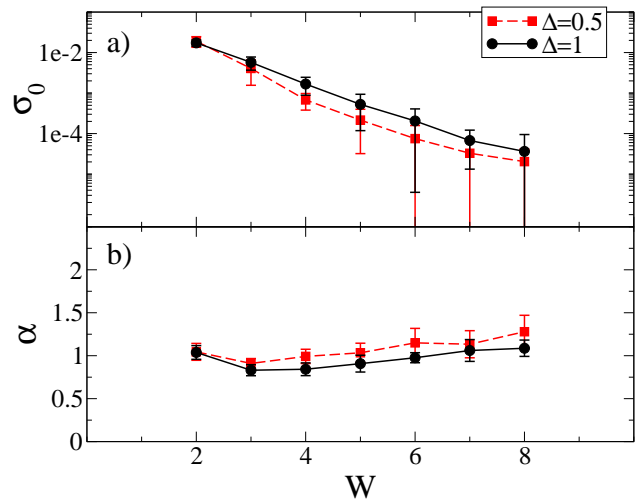}
\caption{(Color online) a) The sample-averaged  d.c. conductivity $\sigma_0$ and the corresponding STS fluctuations, and b)  the sample-averaged low-$\omega$ exponent $\alpha$ and the corresponding STS fluctuations, obtained by the MCLM as functions of disorder $W$ and for $\Delta=0.5,\;1$, $L=28$ sites, $\eta=0.002$. Notice that the bars represent the STS fluctuations, values of which are obtained using the same estimate as for $r_\eta(\omega)$ in Eq.~(\ref{reta}) .}
\label{fig5}
\end{figure}

We now turn to the sample-averaged $\sigma(\omega)$, which is supposed
to be valid macroscopically validity provided that STS fluctuations
discussed in Sec.~\ref{SecFluct} are modest, $r_{\eta}(\omega)
<1$. The quantity of central importance in this context is the
  sample-averaged d.c. value $\sigma_0$, shown as a function of $W$
in Fig.~\ref{fig5}a for $L =28$. In the crossover regime $2 \leq W
\leq 8$,  the two curves for $\Delta = 0.5, 1$, plotted using
  the logarithmic scale in Fig.~\ref{fig5}a, follow qualitatively a linear dependence, meaning that
$\sigma_0 \propto \exp(-c W)$. Furthermore, it is plausible that $c$ increases as the interaction $\Delta$ is decreased, consistent with $\sigma_0=0$ for the NI $\Delta=0$ case.\cite{barisic10} 
While  the sample-averaged $\sigma_0$ apparently behaves smoothly, well within the MBL regime $r_\eta(0)$ becomes very large. That is, because of the large STS fluctuations $r_\eta(0)$ of $\sigma_0$, shown in terms of bars in Fig.~\ref{fig5}a, we may only give an upper bound for $\sigma_0$. 

The low-$\omega$ exponent $\alpha$, given by Eq.~(\ref{sig}), is fitted in Fig.~\ref{fig5}b within the window  $0\leq\omega\leq0.2$. Unlike some other studies,\cite{agarwal15,gopal15} we take into account that $\sigma_0$ can attain finite values and that the frequency resolution is limited by the STS fluctuations. The STS fluctuations of $\alpha$, denoted by the bars in Fig.~\ref{fig5}b, increase with $W$ and $1/\Delta$. Yet, it should be pointed out that the large STS fluctuations of $\sigma_0$ do not affect the behavior of $\alpha$, the latter exhibiting a much weaker STS fluctuations in Fig.~\ref{fig5}. It seems that $\alpha\sim1$ represents a typical behavior in a broad range of $W$ in Fig.~\ref{fig5}b. The trend\cite{gopal15} towards larger $\alpha>1$ appears to be well seen within the MBL regime $W>6$ only.\cite{steinigeweg15}

\begin{figure}[t]
\includegraphics[width=86mm]{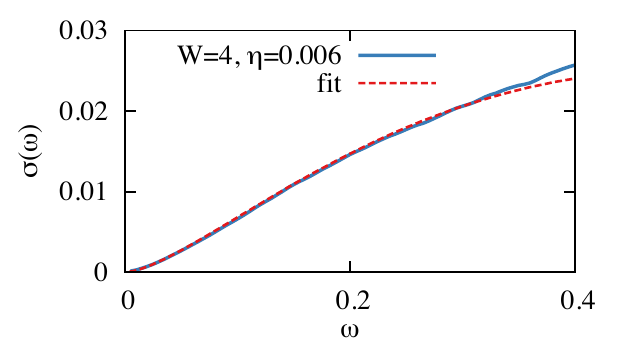}
\caption{(Color online) $\sigma(\omega)$ for the NI ($\Delta=0$) case, evaluated 
for the system of $L=16000$ sites. The fit is given by $\sigma(\omega) \propto \omega^2 \ln^2(b/\omega)$, 
with $b\sim1.37$.}
\label{fig6}
\end{figure}

According to the predictions of the Anderson localization,\cite{mott68,gopal15} one expects that 
$\alpha\rightarrow2$ for $W \gg 1$. In order to show the validity of the latter, we present in Fig.~\ref{fig6} results for the NI system with $W=4$ and $L=16000$ sites. 
Indeed, the numerical data can be described by the NI formula, 
$\sigma_{NI}(\omega) \propto \omega^2\ln^2(b/\omega)$.  However, due to quite large $b>1$, the simplified power law with $\alpha=2$ is restricted to very small $\omega\ll0.1$. In a larger frequency window the same result (obtained for $L\gg100$) can be reasonably fitted with $\alpha=1$. This puts some caveats into the interpretation of fits characterized by $\alpha\sim1$ within the MBL regime as well.

\section{Conclusions}
 
Our results, as shown e.g. in Fig.~\ref{fig1}, reveal that general features of $\sigma(\omega)$  for large disorders 
$W>4 $ are similar irrespective of the interaction (anisotropy) $\Delta$. This suggests that a very short (single-particle) localization length 
$l^* \sim 1$ in this regime suppresses to a large extent the MB effects, induced by $\Delta>0$. The corrections due to $\Delta>0$ are clearly visible in frequency moments, e.g. in $m_2$, Eq.~(\ref{m02}). 
In general, however, the $\Delta>0$ case is highly nontrivial.\cite{huse13,modak15} Still, we can speculate
on the existence of a characteristic frequency scale
$\omega<\Delta \omega$, below which
the interaction $\Delta\neq0$ qualitatively changes $\sigma(\omega)$, whereas  $\Delta \omega$ is  vanishing with increasing $W$ at fixed $\Delta$.

When discussing the STS fluctuations, we should keep in mind that $\sigma(\omega)$ is a global property. Variations of local quantities, 
as e.g. local spin dynamical correlations $S(\omega)$,\cite{herbrych13} may be much larger (involving short system segments), even to the point of a lack of self-averaging in the MBL regime. The relative STS fluctuations in Eq.~(\ref{reta}) should generally scale as 
$1/\sqrt{L}$. With $\omega>0$ fixed, it is plausible that fluctuations become Gaussian in the thermodynamic limit $L\to\infty$.
Nevertheless, $r_{\eta}(\omega)$ are qualitatively different for the NI and the interacting systems. The latter $\Delta>0$ fluctuations
are much smaller, being independent of $\eta$, with the exception of the $\omega\rightarrow0$ limit well 
within the MBL regime $W>6$. This means that even for finite (small) size systems the single-sample $\sigma^k(\omega)$ 
are rather smooth functions, provided that one considers variations beyond the frequency scale set by the level spacing 
$\eta > \delta \epsilon \propto \exp(-\zeta L)$. This property may be qualitatively understood considering the 
frequency moments, given in Eqs.~(\ref{vardel}), and consequently attributed to the correlations in 
$\sigma^k(\omega)$ over the characteristic frequency scale $\Delta \omega$, i.e., over $n_\xi \gg 1$ neighboring frequency bins.  
However, deep in the MBL regime $W \gg 8$, fluctuations increase $r_{\eta}(\omega)>1$  for $L$ considered here. That is, for large $r_{\eta}(\omega)$ at low frequencies $\omega\sim0$, averages may loose their meaning, 
in the analogy to the Griffiths-phase arguments.\cite{gopal15} 

Regarding the transition into the MBL phase, our results seem to favor an interpretation in terms of a crossover rather than a 
qualitative change at a well-defined critical $W=W_c(\Delta)$, although the latter cannot be excluded with our data. 
We observe a continuously exponential-like vanishing of $\sigma_0$ with increasing $W$. This behavior is qualitatively 
compatible with a previous result of the steady increase of $\sigma_0 \propto \Delta$ at fixed $W$ and $\Delta \leq 0.5$, 
derived in the context of the $t$-$V$ model.\cite{barisic10} 
On the other hand, establishing $\sigma_0>0$ deeper in the MBL regime ($W>6$) becomes exceedingly 
difficult due to the STS fluctuations, which restrict the frequency-resolution. 
Yet, either the crossover or the real transition from the ergodic to the MBL regime manifests itself quite sharply
in the very small values of $\sigma_0$ and in the increase of fluctuations.
Therefore, the location of $W_c(\Delta)$ in the phase diagram according to our present results appears quite 
consistent with previous studies.\cite{oganesyan07,monthus10,berkelbach10,lev14,gopal15,steinigeweg15} 
It should be also mentioned that our results for $\sigma_0(W)$ might have a direct relevance 
for experiments on driven fermions on a disordered optical lattice \cite{kondov15}. The steady velocity $v_{cm}$, measured 
in these experiments versus disorder, look very much as our result in Fig.~5a, including the possibility of an interpretation
 of a crossover  rather than a transition at a well defined $W=W_c$ .

Our findings for $\alpha$ in the ergodic regime, $\alpha\sim1$, agrees with some previous and recent studies,\cite{karahalios09,barisic10,steinigeweg15} indicating an anomalous diffusive transport. In some other contexts such a behavior has been traced back to the long-time-tail phenomena related to the nontrivial coupling between hydrodynamic modes.\cite{wilke00} 
However, the origin of such anomalous dynamics in the considered model
remains to be clarified. Moreover, we find $\alpha \sim 1$ even for $W>W_c$.\cite{steinigeweg15} In this latter case,
such $\alpha$ indicates that we are dealing with an insulator characterized by an anomalous dielectric susceptibility
$\chi_0 =\int d \omega \sigma(\omega)/\omega^2$. Within the linear response theory, $\chi_0$ would diverge for $\sigma_0=0$ and $\alpha=1$ (and even faster for $\alpha<1$). Furthermore, such an anomaly within the MBL regime might remain present beyond the linear response approach.\cite{kozarzewski16}
Whatever being the actual case, our results seem to be far from predicted 'normal' insulating behavior $\alpha \sim 2$. We should add a caution that the frequently claimed limiting behavior\cite{mott68,gopal15} is hardly observable even for the NI systems, as shown in Fig.~\ref{fig6}. This is particularly worth noting in the context of expectations that the NI and the interacting $\Delta\neq0$ results merge in the limit $W \gg W_c(\Delta)$.

\begin{acknowledgments}
The authors acknowledge fruitful discussions with R. Steinigeweg,
J. Herbrych and F. Heidrich-Meisner.
This work was supported by the Program P1-0044 of the Slovenian
Research Agency (ARRS) and by the Croatian QuantiXLie Center of Excellence. P.P. acknowledges the support 
by the Alexander von 
Humboldt Foundation.
\end{acknowledgments}

\bibliographystyle{apsrev4-1}
\bibliography{ref_manumbl}

\end{document}